\documentclass[12pt,a4paper]{article}
\usepackage{amsthm,amsfonts,amsmath,amssymb,array,graphicx,tabularx}

\title{An amplitude equation for long, nonlinear internal waves in
a weak shear flow over topography}
\author{M. Yu. Trofimov}
\date{\small\em Il'ichev Pacific Oceanological Institute of the Russian Academy of
Sciences, Vladivostok, 690041, Russia}

\begin{document}
\maketitle
\begin{abstract}
{
A forced, variable coefficients Kor\-te\-weg-de Vries equation for amplitudes
of long, nonlinear internal waves in a stratified shear flow over topography
is derived when the magnitude of the basic flow is small. The derivation is
done by incorporating the basic flow in the perturbation series of the
Taniuti and Wei's reductive perturbation method. Istead of the long-wave
Tailor-Goldstein spectral problem, only
the usual long-wave spectral problem for rested stratified media
and a boundary value problem have to be
solved for calculating the coefficients of obtained equation. Explicit
expressions for these coefficients are presented.
}
\end{abstract}

\noindent
Known amplitude equations of the Kor\-te\-weg-de Vries type for long
nonlinear internal waves in stratified shear flows are based on the long-wave
Tailor-Goldstein spectral problem, which is quad\-ra\-tic in spectral
parameter. Contrary to the ordinary self-adjoint spectral problems arising in
the absence of basic flow, the quadratic ones are hardly treated numerically
because the corresponding algorithms are either complicated or inefficient.
In our opinion, this fact restricts the use of such amplitude equations in
concrete applications. On the other hand, in many interesting applications
the basic flow can be treated as weak. The main idea of our work is to
incorporate this flow in the perturbation series of the Taniuti and Wei's
reductive perturbation method \cite{tan-wei}. The amplitude equation,
obtained in this way, turns out to be based on a first-order perturbation of
the usual long-wave spectral problem for rested stratified media.
This leads to the comparatively simple procedure of calculating the
coefficients of this equation.
\par
We shall use the equations of two-dimensional motions of inviscid,
incompressible stably stratified fluid, written in the form
%
%
\begin{equation} \label{bas1}
\begin{split}
& \rho\left[\partial_t\Delta\Psi+J(\Delta\Psi,\Psi)\right]=
\\
& \frac{1}{\beta}\rho_x-
 \left\{\rho_z\left[\Psi_{tz}+
J(\Psi_z,\Psi)\right]
+\rho_x\left[\Psi_{tx}+J(\Psi_x,\Psi)\right]\right\}
\end{split}
\end{equation}
%
%
\begin{equation} \label{bas2}
\rho_t+J(\rho,\Psi)=0
\end{equation}
%
%
with the boundary conditions
\begin{equation} \label{bc}
 \Psi = 0 \quad \mbox{at} \quad z = 0, \qquad
   \Psi = 0 \quad \mbox{at} \quad z = -H(x)
\end{equation}
where $()_t=\partial_t=\partial /\partial t$, etc.,
$\Delta=\partial^2_x + \partial^2_z$,
$J(\alpha,\beta)=\alpha_x\beta_z - \alpha_z\beta_x$, $\Psi$ is a stream
function, $\rho$ is the density and $H$ describes the bottom topography.
The variables are nondimensional, based on a length scale $\bar h$ (a typical
vertical dimension), a time scale ${\bar N}^{-1}$ (where $\bar N$ is
a typical value of the Brunt-V\"ais\"al\"a frequency), and a density scale
$\bar \rho$ (a typical value of the density). The parameter $\beta$ is
$\bar h {\bar N}^2 g^{-1}$, where $g$ is the gravity acceleration.
\par
For applying the reductive perturbation method to the
equations~(\ref{bas1}), (\ref{bas2}), we introduce a small parameter
$\epsilon$ and slow variables
$\displaystyle \xi = \epsilon\left(\int^{\,x}\frac{1}{c}\,dx' - t\right)$,
$ \eta = \epsilon^3x$,
where
$c = c_0 + \epsilon c_1 + \epsilon^2 c_2 + \ldots$
is the phase velocity of the waves to be considered. The amplitude of these
waves will be $O(\epsilon^2)$ with the amplitude of basic flow
$O(\epsilon)$, so the introduced space-time scales are characteristic for
the Kor\-te\-weg-de Vries theory.
The derivatives with respect to $x$
and $t$ are expressed in new variables as follows:
%
%
\begin{equation} \label{der}
\partial_x = \left(\epsilon\frac{1}{c_0} - \epsilon^2\frac{c_1}{c_0^2} +
\epsilon^3\frac{c_1^2}{c_0^3} - \epsilon^3\frac{c_2}{c_0^2} + \ldots\right)
\partial_\xi + \epsilon^3\partial_\eta\,, \qquad
\partial_t = - \epsilon\partial_\xi
\end{equation}
\par
Expanding $\rho$ and $\Psi$ as $\rho = \rho_0 + \epsilon\rho_1 +
\epsilon^2\rho_2 + \ldots$, $\Psi = \epsilon\Psi_1 + \epsilon^2\Psi_2 +
\ldots$ and substituting these expansions and Eq.~(\ref{der}) into
Eqs.~(\ref{bas1}), (\ref{bas2}) and (\ref{bc}), we obtain a hierarchy of equations
at orders of $\epsilon$. We will search a solution satisfying this
hierarchy up to the order $O(\epsilon^4)$.
In the sequel the bottom topography $H$ is assumed to be dependent only on
$\eta$.
\par
At $O(\epsilon)$ we have $\rho_{0\xi} = 0$, so $\rho_0$ does not depend on
$\xi$.
\par
At $O(\epsilon^2)$  we have
%
%
\begin{eqnarray*}
-\rho_0\Psi_{1zz\xi} & = & \frac{1}{\beta c_0}\rho_{1\xi} +
\rho_{0z}\Psi_{1z\xi} \\
\rho_{1\xi} - \rho_{0z}\frac{1}{c_0}\Psi_{1\xi} & = & 0
\end{eqnarray*}
We suppose that this system is satisfied by the given basic flow which
does not depend on $\xi$. This flow must satisfy the equations (\ref{bas1}),
(\ref{bas2}) up to the order $O(\epsilon^4)$, which gives
%
%
\begin{equation} \label{cond1}
\rho_{0\eta} = 0, \quad \rho_{1\eta} = 0
\end{equation}
\begin{equation} \label{cond2}
\rho_{0z} \Psi_{1\eta} = 0
\end{equation}
By Eq.~(\ref{cond1}), $\rho_0$ and $\rho_1$ both depend only on $z$ and
without loss of generality we take $\rho_1 = 0$. According to
Eq.~(\ref{cond2}), the basic flow has to be pure shear within the pycnocline.
\par
At $O(\epsilon^3)$  we have
%
%
\begin{eqnarray*}
-\rho_0\Psi_{2zz\xi} & = & \frac{1}{\beta c_0}\rho_{2\xi} +
\rho_{0z}\Psi_{2z\xi} \\
\rho_{2\xi} - \rho_{0z}\frac{1}{c_0}\Psi_{2\xi} & = & 0
\end{eqnarray*}
As a solution of these equations we take
%
%
\begin{equation} \label{psi2-rho2}
\Psi_2 = c_0(\eta) A(\eta,\xi) Y(\eta,z)\,, \quad
\rho_2 = - \rho_{0z} A(\eta,\xi) Y(\eta,z)\,,
\end{equation}
where $A$ has the meaning of isopycnal elevation amplitude and $c_0$ together
with $Y$ satisfy the usual long-wave spectral problem (see e.g. Grimshaw et
al. \cite{gr}-\cite{gr-sm})
%
%
\begin{equation} \label{spec}
c_0^2 (\rho_0 Y_z)_z - \frac{1}{\beta}\rho_{0z} Y = 0 \,, \qquad
Y(0) = Y(-H(\eta)) = 0
\end{equation}
with $Y$ normed by
%
%
$$
-\frac{1}{\beta}\int^{\,0}_{-H(\eta)}\rho_{0z} Y^2\,dz = 1
$$
%
\par
At $O(\epsilon^4)$, taking into account Eqs.~(\ref{cond1}) and (\ref{cond2}),
we have
%
%
\begin{eqnarray*}
\lefteqn{%
-\rho_0\Psi_{3zz\xi} + \rho_0\frac{1}{c_0}\Psi_{1z}\Psi_{2zz\xi} -
\rho_0\frac{1}{c_0}\Psi_{2\xi}\Psi_{1zzz}  =
} \\
& &
\frac{1}{\beta}\frac{1}{c_0}\rho_{3\xi} -
\frac{1}{\beta}\frac{c_1}{c_0^2}\rho_{2\xi}
+ \rho_{0z}\Psi_{3z\xi} - \frac{1}{c_0}\rho_{0z}\Psi_{1z}\Psi_{2z\xi} +
\frac{1}{c_0}\rho_{0z}\Psi_{1zz}\Psi_{2\xi}
\\
& & \rho_{3\xi} - \frac{1}{c_0}\rho_{2\xi}\Psi_{1z} +
\frac{1}{c_0}\rho_{0z}\Psi_{3\xi} - \frac{c_1}{c_0^2}\rho_{0z}\Psi_{2\xi}
 = 0
\end{eqnarray*}
Introducing the horizontal velocity $U=\Psi_{1z}$ and postulating an ansatz
for $\Psi_3$ of the form $c_0(\eta) A(\eta,\xi) W(\eta,z)$, after some
calculations we obtain from these equations a boundary value problem for $W$
%
%
\begin{equation} \label{bp_psi3}
\begin{split}
& c_0^2 (\rho_0 W_z)_z - \frac{1}{\beta}\rho_{0z} W =
\\
& 2 c_0 U (\rho_0 Y_z )_z - 2 c_1 c_0 (\rho_0 Y_z )_z
- c_0 \rho_0 U_{zz} Y - c_0 \rho_{0z} U_z Y
\\
&  W(0) = W(-H(\eta)) = 0
\end{split}
\end{equation}
It is remarkable that this boundary value problem coincides with the
$O(\epsilon)$ part of the long-wave Taylor-Goldstein problem
%
%
\begin{eqnarray*}
\lefteqn{%
(\epsilon U - c_0 - \epsilon c_1) \left[ \rho_0 (\epsilon U - c_0 -
\epsilon c_1) (Y + \epsilon W)_z \right]_z -
} & & \\ & &
(\epsilon U - c_0 - \epsilon c_1) \left[ \rho_0 \epsilon U_z (Y + \epsilon W)
\right]_z - \frac{1}{\beta} \rho_{0z} (Y + \epsilon W) = 0
\\ \\ & &
Y(0) + \epsilon W(0) = Y(-H(\eta)) + \epsilon W(-H(\eta)) = 0
\end{eqnarray*}
which is a perturbation of Eq.~(\ref{spec}).
\par
The solvability condition for Eq.~(\ref{bp_psi3}), which consists in
orthogonality of the right-hand side of Eq.~(\ref{bp_psi3}) to $Y$ in
$L^2[-H,0]$, gives the value for the first order correction to the
phase velocity
%
%
\begin{equation} \label{c1}
c_1 = c_0^2\int^{\,0}_{-H}\rho_0 U (Y_z)^2\,dz
\end{equation}
\par
At $O(\epsilon^5)$ we have the system too large to be shown here.
It can be reduced to a boundary value problem for $\Psi_{4\xi}$ of the form
%
%
\begin{eqnarray*}
c_0^2 (\rho_0 \Psi_{4\xi z})_z - \frac{1}{\beta}\rho_{0z} \Psi_{4\xi} =
\mbox{R.-H.~S.}
\\
\Psi_{4\xi}(0) = \Psi_{4\xi}(-H) = 0
\end{eqnarray*}
where R.-H.~S. stands for a collection of terms containing $A$, $Y$, $W$,
$c_0$ and $c_1$.
The solvability condition for this problem, which is the same as for
Eq.~(\ref{bp_psi3}), gives an amplitude equation of the Kor\-te\-weg-de Vries
type
%
%
\begin{equation} \label{kdv}
\alpha A + A_\eta + \beta A A_\xi + \gamma A_\xi +
\delta  A_{\xi \xi \xi} = {\cal F}
\end{equation}
where
%
%
\begin{eqnarray*}
\alpha  & = & \frac{1}{2} \frac{\partial}{\partial \eta} \ln c_0 \\
 \\
\beta   & = & \frac{3}{2} c_0 \int^{\,0}_{-H} \rho_0 (Y_z)^3\,dz \\
 \\
\gamma  & = &
- 2 \int^{\,0}_{-H} \rho_0 (U - c_1) W_z Y_z \, dz -
  \int^{\,0}_{-H} \rho_0 U_z (W_z Y - W Y_z) \, dz +
 \\
& &
\frac{1}{c_0} \int^{\,0}_{-H} \rho_0 \left\{( U - c_1 )^2 \left[Y_z\right]^2
 - \left[( U - c_1 )_z\right]^2 Y^2 \right\} \, dz \\
 \\
\delta  & = & \frac{1}{2 c_0} \int^{\,0}_{-H} \rho_0 Y^2\,dz \\
 \\
{\cal F} & = & \int^{\,0}_{-H} \rho_0 \left( U^2 \right)_{\eta} Y_z \, dz +
\rho_0(-H) [ U(-H)]^2 Y_z(-H) H_{\eta}
\end{eqnarray*}
It should be noted that only $\gamma$ and ${\cal F}$ are influenced by the
basic flow. The expressions for other coefficients are fully equivalent
to the expressions for the corresponding coefficients of the extended
Kor\-te\-weg-de Vries equation, used by Helfrich and Melville \cite{hel-mel},
where the basic flow were not considered.
Note also that the forcing term of the amplitude equation derived by
Grimshaw and Smyth \cite{gr-sm} is just a particular case of our ${\cal F}$.
\par
We conclude that for the evaluation of the coefficients of the equation
(\ref{kdv}) only the spectral problem (\ref{spec}) and the boundary problem
(\ref{bp_psi3}) have to be solved. The first problem is a standard one and
its numerical treatment is well developed. The second is not so standard
because of its degeneracy, but it has a unique solution in the class of
functions orthogonal to $Y$ in $L^2[-H,0]$. This solution can be found
numerically by the special choice of basis or by other known methods. As it
is easily checked, any other solution produces the same value of the
coefficients $\gamma$ and ${\cal F}$.

\end{document}